\documentclass[12pt,english]{iopart}
\usepackage{cite}
\usepackage{graphicx}
\usepackage{iopams}
\usepackage{bm}

\usepackage[usenames,dvipsnames,svgnames,table]{xcolor}
\usepackage[unicode=true,
            pdfusetitle,
            bookmarks=true,
            bookmarksnumbered=false,
            bookmarksopen=false,
            breaklinks=true,
            pdfborder={0 0 0},
            backref=false,
            colorlinks=true]{hyperref}
\hypersetup{linkcolor=NavyBlue,urlcolor=NavyBlue,citecolor=NavyBlue}

\providecommand{\openone}{\leavevmode\hbox{\small1\kern-4.3pt
\normalsize1}}

\begin{document}

\title{Quantum Fisher information and symmetric logarithmic derivative via anti-commutators}

\author{Jing Liu}
\address{State Key Laboratory of Modern Optical Instrumentation, 
Department of Physics, Zhejiang University, Hangzhou 310027, China}
\address{Department of Mechanical and Automation Engineering, 
The Chinese University of Hong Kong, Shatin, Hong Kong}

\author{Jie Chen}
\address{State Key Laboratory of Modern Optical Instrumentation, 
Department of Physics, Zhejiang University, Hangzhou 310027, China}

\author{Xiao-Xing Jing}
\address{State Key Laboratory of Modern Optical Instrumentation, 
Department of Physics, Zhejiang University, Hangzhou 310027, China}

\author{Xiaoguang Wang}
\address{State Key Laboratory of Modern Optical Instrumentation, 
Department of Physics, Zhejiang University, Hangzhou 310027, China}
\address{Synergetic Innovation Center of Quantum Information  
and Quantum Physics, University of Science and Technology of China, 
Hefei, Anhui 230026, China}
\ead{xgwang@zimp.zju.edu.cn}

\begin{abstract}
Symmetric logarithmic derivative (SLD) is a key quantity to obtain
quantum Fisher information (QFI) and to construct the corresponding
optimal measurements. Here we develop a method to
calculate the SLD and QFI via anti-commutators. This method is originated from the 
Lyapunov representation and would be very useful for cases that the anti-commutators 
among the state and its partial derivative exhibits periodic properties. As an application, 
we discuss a class of states, whose squares linearly depend on the states themselves, 
and give the corresponding analytical expressions of SLD and QFI. A noisy scenario of 
this class of states is also considered and discussed. Finally, we readily apply the method 
to the block-diagonal states and the multi-parameter estimation scenarios.
\end{abstract}

\pacs{03.67.-a, 03.65.Ta, 06.20.-f}

\maketitle

\section{Introduction}
Quantum metrology has been going through a great development in recent
years~\cite{intro1, intro2, intro3, intro4, intro5, intro6, intro7, intro8,
intro9,intro9_1,intro10,intro11,LiuPRA,intro12,intro13,Salvatori}.
Quantum Fisher information (QFI) is a crucial concept in quantum metrology as it 
depicts the lower bound on the variance of an unbiased estimator
for the parameter under estimation, according to the quantum Cram\'{e}r-Rao 
theorem~\cite{Helstrom,Holevo}. The definition of QFI is 
$F:=\langle L^{2}\rangle=\mathrm{Tr}(\rho L^{2})$~\cite{Helstrom,Holevo},
where $L$ is so-called symmetric logarithmic derivative (SLD). Denoting
the parameter under estimation as $\theta$, the SLD operator is determined by the equation
\begin{equation}
\partial_{\theta}\rho=\frac{1}{2}\left(\rho L+L\rho\right).\label{eq:SLD}
\end{equation}
Taking the trace on both sides of this equation, one can see that
$\langle L\rangle=0$. Therefore, the QFI is actually the variance of SLD operator, i.e., 
$F=\langle\Delta^{2}L\rangle$, with $\Delta^{2}L:=(L-\langle L\rangle)^{2}$.

The SLD operator has been studied for years~\cite{Braunstein,Braunstein1,TothJPA,
Liu1,Liu2,Zhang,Marzolino,Monras,Paris,Ercolessi}. 
It is important for two reasons. First, it is obvious
that the QFI can be directly obtained when the SLD operator is known. 
Second, the achievement of quantum Cram\'{e}r-Rao bound strongly 
depends on the measurement, namely, it can only be achieved for some 
optimal measurements. The eigenvectors of SLD operator are such 
theoretical optimal measurements~\cite{Braunstein,Braunstein1,TothJPA} 
if they are locally independent of the parameter.  
Thus, the study of SLD operator could help us to construct or find
optimal measurements for the achievement of the highest precision.

The traditional method for the calculation of SLD operator is to expand it in the eigenspace 
of density matrix. We now denote the spectral decomposition of 
$\rho$ as $\sum_{i=1}^{M}p_{i}|\psi_{i}\rangle\langle\psi_{i}|$,
with $p_{i}$, $|\psi_{i}\rangle$ the $i$th eigenvalue and eigenstate
of $\rho$, respectively. $M$ is the dimension of $\rho$'s support. When $\rho$ is positive definite (or full rank), 
$M$ equals to the state's dimension $d$. Utilizing the spectral decomposition form, 
the element of SLD operator can be expressed by~\cite{Liu1,Liu2,Zhang,Paris}
\begin{equation}
L_{ij}=\frac{\partial_{\theta}p_{i}}{p_{i}}\delta_{ij}
+\frac{2(p_{i}-p_{j})}{p_{i}+p_{j}}\langle\partial_{\theta}\psi_{i}
|\psi_{j}\rangle,
\end{equation}
for $i,j\in[1,M]$. $L_{ij}$ can be an arbitrary
number when either $i$ and $j$ is larger than $M$. This method to
calculate SLD operator is useful when the spectral decomposition of
$\rho$ is not difficult to obtain. However, for many cases, for example 
when the space of the density matrix is infinitely large, the diagonalization 
could be very tricky.

Lyapunov representation is another method to obtain the SLD operator
and applied in many scenarios~\cite{Marzolino,Monras,Paris}.
The definition equation~(\ref{eq:SLD}) is actually a special
form of Lyapunov equation, indicating that SLD operator is
a corresponding solution. In this representation, the SLD operator
is expressed by~\cite{Paris}
\begin{equation}
L=2\int_{0}^{\infty}e^{-\rho s}\left(\partial_{\theta}\rho\right)e^{-\rho s}ds.\label{eq:Lyapunov_L}
\end{equation}
One advantage of this representation is that it is basis-independent.
Generally, this representation is no easier to calculate
than the traditional method. However, similar to the traditional one,
Lyapunov representation would be very useful for some scenarios. In
this paper, we first review the Lyapunov representation and figure
out that Eq.~(\ref{eq:Lyapunov_L}) is available for both full and non-full 
rank density matrices. Then we provide a further basis-independent expression 
of SLD operator based on the Lyapunov form. The new expression would be 
extremely useful when the anti-commutator between the density matrix 
and its partial derivative exhibits periodic properties.

To show the advantage of this method, we apply it in a class of
states showing a linear relation with their squares. This class
includes all pure and all two-level states. We provide simple expressions
of SLD operator and corresponding QFI via the given method for this class.  
Noise from the environment are widely exist in reality. The scenario for these 
states under depolarizing channel are considered. Moreover, we also discuss the block 
diagonal states and the multiparameter estimations.

\section{Lyapunov representation}

As the beginning, we first review the derivation of Lyapunov representation of SLD operator. 
Mathematically, Eq.~(\ref{eq:SLD}) is  known as a special form
of Lyapunov equation. To solve this equation, one can construct a
function
\begin{equation}
f(s)=e^{-\rho s}Le^{-\rho s}, \label{eq:def}
\end{equation}
which satisfies $f(0)=L$.
The partial derivative of $f(s)$ on $s$ is $\partial_{s}f(s)=-2e^{-\rho s}
\left(\partial_{\theta}\rho\right)e^{-\rho s}.$
Integrating both sides of this derivative equation, one can obtain
\begin{equation}
f(\infty)-f(0)=-2\int_{0}^{\infty}e^{-\rho s}(\partial_{\theta}\rho)
e^{-\rho s}ds \label{eq:ffff}.
\end{equation}
When $\rho$ is full rank, $e^{-\rho s}$ trends to zero for
$s\rightarrow\infty$, indicating that $f(\infty)=0$.
Thus, the SLD operator can be directly written in the form of Eq.~(\ref{eq:Lyapunov_L}).
However, when $\rho$ is non-full rank, $f(\infty)$ will not
vanish. Reminding that $M$ and $d$ are the dimensions of $\rho$'s support
and $\rho$ respectively, then the limitation of $e^{-\rho s}$
equals to $\mathrm{diag}\left\{ 0_{M},\openone_{d-M}\right\} $ when
$s$ trends to positive infinite. Here $0_{M}$ is the $M$-dimensional
zero matrix and $\openone_{d-M}$ is the $(d-M)$-dimensional identity
matrix. Correspondingly, we manually separate the SLD operator
into four blocks as
\begin{equation}
L=\left(\begin{array}{cc}
A_{M} & B_{d-M,M}\\
B_{d-M,M}^{\dagger} & C_{d-M}
\end{array}\right),
\end{equation}
where the Hermiticity of $L$ is applied. Utilizing this form, and
based on Eq.~(\ref{eq:def}), one can see that $f(\infty)=\mathrm{diag}\left\{ 0_{M},C_{d-M}\right\} .$

Meanwhile,  on the right side of Eq.~(\ref{eq:ffff}),  in the eigenbasis of $\rho$,
$e^{-\rho s}$ is block diagonal, and the element $[\partial_{\theta}\rho]_{ij}$
vanishes for $i>M$ and $j>M$~\cite{Liu2}. Then after the integral, the elements of the entirety 
for $i,j\in[d-M,d]$ keep vanishing. This means the block $C_{d-M}$ is not involved in 
Eq.~(\ref{eq:ffff}), and cannot be solved by this equation. Thus, $C_{d-M}$ is actually 
undefined here. However, since $C_{d-M}$ is also not involved in the calculation of 
QFI~\cite{Liu1,Liu2}, therefore, it will not bring indeterminacy on the final expression 
of QFI. Based on this reason, we can simply take $C_{d-M}=0$ for convenience. In this 
way, the SLD operators for both full and non-full rank density operators can be uniformly 
expressed in the form of  Eq.~(\ref{eq:Lyapunov_L}).

\subsection{Expanded representation} 

Defining the anti-commutator $\rho^{\mathrm{o}}$ as 
$\rho^{\mathrm{o}}(\cdot)=\{\rho,\cdot\}$ and noticing the
fact that the improper integral $\int_{0}^{\infty}$ can also be written
as ${\mathrm{lim}}_{s\rightarrow\infty}\int_{0}^{s}$, Eq.~(\ref{eq:Lyapunov_L})
can be further expressed in the form
\begin{equation}
L=-2\lim_{s\rightarrow\infty}\sum_{n=0}^{\infty}
\frac{\left(-s\right)^{n+1}}{(n+1)!}
\left(\rho^{\mathrm{o}}\right)^{n}\partial_{\theta}\rho.
\label{eq:L}
\end{equation}

This is a further basis-independent form of SLD. This
formula would be very useful when the anti-commutator among 
$\rho$ and its partial derivative exhibits periodic properties. In some cases, 
$\partial_{\theta}\rho$ is the eigen-operator of $\rho^{\mathrm{o}}$, i.e., 
$\rho^{\mathrm{o}}\partial_{\theta}\rho=a\partial_{\theta}\rho$
with $a$ a real number. When $a>0$, the SLD operator
reduces to a very simple form
\begin{equation}
L=\frac{2}{a}\partial_{\theta}\rho.
\end{equation}
The simplest case here is the pure states. For a pure state, it is easy
to see that $a=1$ as $\rho^{2}=\rho$. Thus, the SLD operator for
pure states is $L=2\partial_{\theta}\rho$.

Moreover, because of the equality
\begin{equation}
\left(\rho^{\mathrm{o}}\right)^{n}\partial_{\theta}\rho=\sum_{m=0}^{n}
C_{n}^{m}\rho^{m}\left(\partial_{\theta}\rho\right)\rho^{n-m},
\end{equation}
where $C_{n}^{m}=n!/[m!(n-m)!]$, the SLD operator in Eq.~(\ref{eq:L})
can also be written in an equivalent form
\begin{equation}
L=-2\lim_{s\rightarrow\infty}\sum_{n=0}^{\infty}\sum_{m=0}^{n}
\frac{\left(-s\right)^{n+1}}
{(n+1)!}C_{n}^{m}\rho^{m}\left(\partial_{\theta}\rho\right)
\rho^{n-m}.\label{eq:L_C}
\end{equation}
This form of SLD operator could be very useful when $\rho^{m}\left(\partial_{\theta}\rho\right)\rho^{n-m}$
are easy to calculate. Especially, when $\rho$ commutes with $\partial_{\theta}\rho$,
above formula reduces to~\cite{Paris}
\begin{equation}
L=\rho^{-1}\partial_{\theta}\rho=\left(\partial_{\theta}\rho\right)\rho^{-1},
\label{eq:L_commute}
\end{equation}
where the equality $\sum_{m=0}^{n}C_{n}^{m}=2^{n}$ has been applied.
The detailed calculation can be found in the appendix.
Based on this equation, the corresponding QFI reads
\begin{equation}
F=\mathrm{Tr}[\rho^{-1}(\partial_{\theta}\rho)^{2}].
\end{equation}
One vivid example here is the parameter only shows up in the eigenvalues
of $\rho$. In this case, $\partial_{\theta}\rho$ and $\rho^{-1}$ are both 
diagonal in the eigenspace of $\rho$. Thus, the SLD operator is in the
form of Eq.~(\ref{eq:L_commute}). More specifically, in the eigenspace 
of $\rho$, it is $L_{ij}=\delta_{ij}\partial_{\theta}\ln p_{i}.$

Recently, Correa \emph{et al.}~\cite{Correa} studied a quantum thermometer 
prototype and demonstrated that after a full thermalization, the optimal probe 
state to access the maximum QFI is an effective two-level system with a highly 
degenerate excited state, of which the Hamiltonian can be
written as $H=\lambda_{1}(\beta)\sum^{d-1}_{i=1}|\xi_i\rangle\langle\xi_i|
+\lambda_2(\beta)|\xi_d\rangle\langle\xi_d|$.
$\lambda_1$ and $\lambda_2$ here are temperature dependent
energy gaps and $|\xi_i\rangle$ is the eigenstate of Hamiltonian for any $i$. 
In this case, since $|\xi_i\rangle$ is temperature independent,
the thermal state $\rho=\exp(-\beta H)/Z$ is commutative with
its partial derivative on temperature. Thus, its SLD operator can be expressed 
in Eq.~(\ref{eq:L_commute}). Specifically, it is 
$L=-(\lambda_1+\beta\partial_{\beta}\lambda_1)\sum^{d}_{i=1}
|\xi_{i}\rangle\langle\xi_{i}|-(\lambda_2+\beta\partial_{\beta}\lambda_2)
|\xi_{d}\rangle\langle\xi_{d}|.$ Substituting the expressions of 
$\lambda_1$ and $\lambda_2$ in this equation, the QFI in Ref.~\cite{Correa} can be reproduced.

\subsection{Unitary parametrization}

A unitary parametrization process contains a large category of realistic parametrization processes.
Recently, an alternative representation of QFI for unitary parametrization processes 
has been discussed~\cite{Pang,Liu3,Skotiniotis}.
For a unitary parametrization process, the parametrized state 
$\rho=U(\theta)\rho_{\mathrm{in}}U^{\dagger}(\theta)$,
with $U(\theta)$ a $\theta$-dependent unitary matrix. The initial
state $\rho_{\mathrm{in}}$ is independent of $\theta$.
A key quantity in the alternative
representation is a Hermitian operator $\mathcal{H}=i(\partial_{\theta}U^{\dagger})U$.
All the information of parametrization is involved in this basis-independent
operator. For a unitary process, the QFI can be expressed by~\cite{Paris,Liu3}
\begin{equation}
F=\mathrm{Tr}\left(\rho_{\mathrm{in}}L_{\mathrm{eff}}^{2}\right),
\end{equation}
where $L_{\mathrm{eff}}=U^{\dagger}LU$ is the effective SLD operator.
In this scenario, it is easy to find that $\rho^{m}\left(\partial_{\theta}\rho\right)
\rho^{n-m}=iU\rho_{\mathrm{in}}^{m}
\left[\mathcal{H},\rho_{\mathrm{in}}\right]\rho^{n-m}U^{\dagger}.$
Based on Eq.~(\ref{eq:L_C}), the effective SLD operator can be written as
\begin{equation}
L_{\mathrm{eff}}=-i2\lim_{s\rightarrow\infty}\sum_{n=0}^{\infty}\sum_{m=0}^{n}
\frac{\left(-s\right)^{n+1}}{(n+1)!}C_{n}^{m}\rho_{\mathrm{in}}^{m}
\left[\mathcal{H},\rho_{\mathrm{in}}\right]\rho_{\mathrm{in}}^{n-m}.
\end{equation}
For a pure initial state, as $\rho_{\mathrm{in}}[\mathcal{H},\rho_{\mathrm{in}}]\rho_{\mathrm{in}}=0$,
$L_{\mathrm{eff}}$ reduces to the known form  $L_{\mathrm{eff}}=i2[\mathcal{H},\rho_{\mathrm{in}}]$~\cite{Liu3}.
When the equation $\left\{ \mathcal{H},\rho_{\mathrm{in}}^{2}\right\} =2\rho_{\mathrm{in}}\mathcal{H}\rho_{\mathrm{in}}$
is satisfied, $\rho_{\mathrm{in}}$ and $[\mathcal{H},\rho_{\mathrm{in}}]$
are commutative. Then the effective SLD operator is $L_{\mathrm{eff}}=i\left(\mathcal{H}-\rho_{\mathrm{in}}\mathcal{H}
\rho_{\mathrm{in}}^{-1}\right).$

\section{Application}
Now we apply Eq.~(\ref{eq:L})
into a class of density operators, which share a common feature
as below
\begin{equation}
\rho^{2}=\alpha\rho-\beta,\label{eq:rho_form}
\end{equation}
where $\alpha$ and $\beta$ are real numbers. In the eigenbasis of
$\rho$, above equation is equivalent to $p_{i}^{2}=\alpha p_{i}-\beta$
for any $i$, which gives the solution $p_{i}=[\alpha\pm\sqrt{\alpha^{2}-4\beta}]/2$.
If only one of the solutions is positive, the density matrix is
trivially proportional to the identity matrix. Thus, we only consider
the situation that both solutions are positive, i.e., $\alpha>2\beta>0$. 
Moreover, it is worth to notice that $\alpha$, $\beta$
in Eq.~(\ref{eq:rho_form}) can either depends on $\theta$ or not.
Several well-known states, including all pure and all two-level
states, satisfy this relation. From Eq.~(\ref{eq:rho_form}),
it is easy to see that $\mathcal{P}=\mathrm{Tr}\rho^{2}=\alpha-\beta$ 
is the purity of $\rho$ and satisfies $d^{-1}\leq\mathcal{P}\leq1$ with $d$ 
the dimension of $\rho$. From Eq.~(\ref{eq:L}), the SLD operator for
this class of states can be expressed by
\begin{equation}
L=\frac{1}{\alpha}\left[2\partial_{\theta}\rho+\left(\partial_{\theta}
\beta\right)\rho^{-1}-\partial_{\theta}\alpha \bm{I}_{\mathrm{M}}\right],
\label{eq:L_category_temp}
\end{equation}
where $\bm{I}_{\mathrm{M}}=\mathrm{diag}\{\openone_{\mathrm{M}},0_{\mathrm{d-M}}\}$ 
is the identity matrix on the support of $\rho$.  The detailed  derivation of this equation 
can be found in the appendix.  If $\rho$ is full-rank, 
$\bm{I}_{M}$ is the identity matrix in the total $d$-dimensional space, 
i.e., $\openone_{\mathrm{d}}$. 
When $\rho$ is non-full rank, i.e., $\det\rho=0$, $\rho^{-1}$
is the inverse matrix of $\rho$ on the support.  Based on previous discussions, 
we know that the elements of SLD outside the support are undetermined and 
they will not affect the value of QFI. Thus, for both full and non-full rank cases, 
$\bm{I}_{\mathrm{M}}$ can be replaced by $\openone_{\mathrm{d}}$, which 
means Ea.~(\ref{eq:L_category_temp}) can be simplified into 
\begin{equation}
L=\frac{1}{\alpha}\left[2\partial_{\theta}\rho+\left(\partial_{\theta}
\beta\right)\rho^{-1}-\partial_{\theta}\alpha\right].
\label{eq:L_category}
\end{equation}

When $\alpha$, $\beta$ are both constant numbers independent of $\theta$,
the SLD operator reduces to $L=2\partial_{\theta}\rho/\alpha$.
Alternatively, $\alpha$ can be a constant and $\beta$ is dependent
on $\theta$. Under this situation, the SLD operator reduces to 
$L=\left[2\partial_{\theta}\rho+(\partial_{\theta}\beta)\rho^{-1}\right]
/\alpha$. A well-known case here is the two-level states, 
which can be expressed in the Bloch representation 
$\rho=\left(\openone_2+\bm{r}\cdot\bm{\sigma}\right)/2$.
Here $\bm{\sigma}=(\sigma_{x},\sigma_{y},\sigma_{z})^{\mathrm{T}}$
is the vector of Pauli matrices and $\bm{r}$ is the Bloch vector.
Utilizing this representation, one can immediately find that $\rho$ satisfies
Eq.~(\ref{eq:rho_form}) with $\alpha=1$ and $\beta=(1-\mathcal{P})/2$.
Here $\mathcal{P}$ is the purity. Then the SLD operator is
\begin{equation}
L=2 \partial_{\theta}\rho-\frac{1}{2}\left(\partial_{\theta}
\mathcal{P}\right)\rho^{-1}. \label{eq:L_qubit}
\end{equation}
This is the general basis-independent expression of SLD operator for
any two-level state. For a mixed state, the inverse of $\rho$ can be written as
\begin{equation}
\rho^{-1}=\frac{2}{1-\mathcal{P}}\sigma_{y}\rho^{\mathrm{T}}\sigma_{y}.
\end{equation}
Thus, for a mixed state, Eq.~(\ref{eq:L_qubit}) can be rewritten into
\begin{equation}
L=2 \partial_{\theta}\rho-\frac{\partial_{\theta}
\mathcal{P}}{1-\mathcal{P}}\sigma_{y}\rho^{\mathrm{T}}\sigma_{y} .
\end{equation}
Substituting the Bloch representation of $\rho$ into above equation,
the entire Bloch representation
of SLD operator in Ref.~\cite{wzhong} can be reproduced.

Equation~(\ref{eq:L_category}) is the general expression of SLD
operator for all states satisfying Eq.~(\ref{eq:rho_form}). Utilizing
this formula, the corresponding basis-independent expression of QFI reads
\begin{eqnarray}
F &=& \frac{1}{\alpha^{2}}\Big[2\alpha\mathrm{Tr}
\left(\partial_{\theta}\rho\right)^{2}
+\left(\partial_{\theta}\beta\right)^{2}\mathrm{Tr}\rho^{-1}
-\left(2M-1\right)\left(\partial_{\theta}\alpha\right)
\partial_{\theta}\beta\Big].\label{eq:F}
\end{eqnarray}
The advantage of this basis-independent expression shows in two aspects. 
First, during the specific calculation, one can choose a convenient basis, 
in which $\partial_{\theta}\rho$ or $\rho^{-1}$ are easy to express or calculate.
Second, utilizing a basis-independent formula, the effects of the
density matrix on the QFI are more distinct.

The fact that Harmonic mean is less than the Arithmetic mean
implies the inequality $\mathrm{Tr}\rho^{-1}\geq M^{2}$. Meanwhile, it
is easy to find $\mathrm{Tr}(\partial_{\theta}\rho)^{2}\geq0$. Thus,
the QFI in Eq.~(\ref{eq:F}) is bounded by the inequality
\begin{equation}
F\geq\frac{1}{\alpha^{2}}\left[\left(M\partial_{\theta}\beta\right)^{2}
-\left(2M-1\right)\left(\partial_{\theta}\alpha\right)
\partial_{\theta}\beta\right].
\end{equation}
This lower bound only depends on the coefficients and can be used
to roughly evaluate the QFI.

For the cases that $\alpha$ and $\beta$ are both $\theta$-independent,
the QFI reduces to $F=2\mathrm{Tr}(\partial_{\theta}\rho)^{2}/\alpha.$
When $\alpha$ is a constant and $\beta$ is dependent
on $\theta$, the QFI is in the form
\begin{equation}
F=\frac{1}{\alpha^{2}}\left[2\alpha\mathrm{Tr}\left(\partial_{\theta}\rho\right)^{2}
+\left(\partial_{\theta}\beta\right)^{2}\mathrm{Tr}\rho^{-1}\right].
\end{equation}
From this equation, the QFI for any two-level state can be immediately obtained, 
which coincides with the equations given in Ref.~\cite{Dittmann}. 
For any mixed two-level state, the QFI is bound by the inequality  
$F_{\mathrm{q}}\geq(\partial_{\theta}\mathcal{P})^{2}$.

\subsection{States under depolarizing channel} 

Most quantum states has to face the disturbance from the environment in reality. 
Depolarizing channel is very usual in quantum processes. Several quantum metrological 
problems of states under depolarizing channel have been discussed recently~\cite{Yao,Xiao,Yao1}. 
In this channel, the final state $\rho_{\mathrm{f}}$ can be expressed by~\cite{Nielsen}
\begin{equation}
\rho_{\mathrm{f}}=\eta\rho_{\mathrm{in}}+\frac{1-\eta}{d}\openone_{d},
\end{equation}
where $\rho_{\mathrm{in}}$ is the initial state and $\eta$ is the
reliability of the channel.

Now consider the situation that the initial state $\rho_{\mathrm{in}}$
satisfies Eq.~(\ref{eq:rho_form}), i.e., $\rho_{\mathrm{in}}^{2}=\alpha_{\mathrm{in}}\rho_{\mathrm{in}}
-\beta_{\mathrm{in}}.$
Under this situation, the final state also satisfies Eq.~(\ref{eq:rho_form})
with the coefficients $\alpha=\eta\alpha_{\mathrm{in}}+2(1-\eta)/d$,
and $\beta=\eta^{2}\beta_{\mathrm{in}}+\eta(1-\eta)\alpha_{\mathrm{in}}/d
+(1-\eta)^{2}/d^{2}$.
In this way, the SLD operator for the final state can be directly
obtained by substituting the specific formula of $\alpha$ and $\beta$
into Eq.~(\ref{eq:L_category}).

If $\alpha_{\mathrm{in}}$ and $\beta_{\mathrm{in}}$ are both $\theta$-independent, 
so will $\alpha$ and $\beta$, the SLD
operator then reads
\begin{equation}
L_{\theta}=\frac{2d\eta}{d\eta\alpha_{\mathrm{in}}+2(1-\eta)}
\partial_{\theta}\rho_{\mathrm{in}}.
\end{equation}
One example for this case is a degenerate mixed state, i.e., 
$\rho_{\mathrm{in}}=\sum_{i=1}^{N}\frac{1}{N}|\psi_{i}(\theta)
\rangle\langle\psi_{i}(\theta)|,$
where $N$ ($N<d$) is the degeneracy and $\langle\psi_{i}(\theta)|\psi_{j}(\theta)\rangle=\delta_{ij}$.
It is easy to see that $\rho_{\mathrm{in}}^{2}=\rho_{\mathrm{in}}/N$,
satisfying Eq.~(\ref{eq:rho_form}). Taking $\theta$ as the parameter
under estimation, the SLD operator can be directly written as
\begin{equation}
L_{\theta}=\frac{d\eta}{d\eta+2N(1-\eta)}\sum_{i=1}^{d}L_{\mathrm{in},i},
\end{equation}
where $L_{\mathrm{in},i}=2\partial_{\theta}(|\psi_{i}\rangle\langle\psi_{i}|)$
is the SLD operator for $|\psi_{i}\rangle\langle\psi_{i}|$. For a pure initial
state, i.e., $N=1$, the SLD operator reduces to the known form discussed
in Ref.~\cite{Yao1}.

Alternatively, the reliability $\eta$ can also be the parameter
under estimation. In this case, it can be checked that $\rho_{\mathrm{f}}$
commutes with $\partial_{\eta}\rho_{\mathrm{f}}$. Thus, for any
form of input state $\rho_{\mathrm{in}}$, the SLD operator here is always
in the form of Eq.~(\ref{eq:L_commute}), namely, 
$L_{\eta}=\rho_{\mathrm{f}}^{-1}\partial_{\eta}\rho_{\mathrm{f}}$. Denoting the 
spectrol decomposition of $\rho_{\mathrm{in}}$ as 
$\sum_{i}\lambda_{i}|\lambda_{i}\rangle\langle\lambda_{i}|$, 
the SLD operator can be expressed by 
\begin{equation}
L_{\eta}=\sum_{i}\frac{d\lambda_{i}-1}{\eta(d\lambda_{i}-1)+1}|\lambda_{i}\rangle\langle\lambda_{i}|.
\end{equation}
From above equation, one can see that the eigenbasis of initial state could be the optimal measurements 
to access the quantum Fisher information.  The corresponding QFI is in then in the form
\begin{equation}
F_{\eta}=\sum_{i}\frac{(d\lambda_{i}-1)^2}{d[\eta(d\lambda_{i}-1)+1]}.
\end{equation}
If the initial state is a pure state, i.e., 
$|\lambda\rangle\langle\lambda|$, the SLD reduces to 
\begin{equation}
L_{\eta} = \frac{d}{(1-\eta)(\eta d+1-\eta)}|\lambda\rangle\langle\lambda|
-\frac{1}{1-\eta}\openone_{d},
\end{equation}
and the QFI is 
\begin{equation}
F_{\eta}=\frac{(d-1)^2}{(d\eta+1-\eta)^2}.
\end{equation}
From this equation, it can be found that with the increase of $\eta$, 
the value of $F_{\eta}$ monotonously reduces.

\subsection{Block diagonal states}

The block diagonal states are widely used and discussed
in quantum mechanics. One vivid example is the optical systems
taking into account the superselection rules~\cite{Hyllus}. Generally,
a block diagonal state can be written as $\rho=\bigoplus_{i=1}^{n}\rho_{i}.$
Here $\bigoplus$ represents the direct sum. One can check that
an available form of SLD operator here is block diagonal, i.e., $L=\bigoplus_{i=1}^{n}L_{i}$,
where $L_{i}$ is the corresponding SLD operator for $\rho_{i}$. Consider
the scenario that each block satisfies the equation $\rho_{i}^{2}=\alpha_{i}\rho_{i}-\beta_{i}.$
It should be noticed that $\mathrm{Tr}\rho_{i}<1$ and the purity
for $\rho$ is $\mathcal{P}=\sum_{i}\alpha_{i}\mathrm{Tr}\rho_{i}-\beta_{i}$.
In this scenario, each $L_{i}$ satisfies Eq.~(\ref{eq:L_category}).
Thus, the entire SLD operator can be expressed by
\begin{equation}
L=\bigoplus_{i=1}^{n}\frac{1}{\alpha_{i}}\left[2\partial_{\theta}\rho_{i}
+\left(\partial_{\theta}\beta_{i}\right)\rho_{i}^{-1}
-\partial_{\theta}\alpha_{i}\right].
\end{equation}

Moreover, if $\rho_{i}$ is a 2-dimensional block, it can be expanded
via the Pauli matrices into $\rho_{i}=\mu_{i}\openone+\bm{r}_{i}\cdot\bm{\sigma},$
where $\mu_{i}=\mathrm{Tr}\rho_{i}/2$, and $\bm{r}_{i}$ is the Bloch
vector for $i$th block. Then it can be found that $\rho_{i}$ satisfies Eq.~(\ref{eq:rho_form})
with $\alpha_{i}=2\mu_{i}$ and $\beta_{i}=2\mu_{i}^{2}-\mathcal{P}_{i}/2$.
Here $\mathcal{P}_{i}=\mathrm{Tr}\rho_{i}^{2}$. With these coefficients,
the SLD operator for $i$th block reads
\begin{equation}
L_{i}=\frac{1}{\mu_{i}}\left(\partial_{\theta}\rho_{i}+\xi_{i}\rho_{i}^{-1}
-\partial_{\theta}\mu_{i}\right),
\end{equation}
where the coefficient $\xi_{i}=2\mu_{i}\partial_{\theta}\mu_{i}-\partial_{\theta}\mathcal{P}_{i}/4$.
If $\det\rho_{i}=0$, $\xi_{i}$ vanishes. All the X states can fit in this scenario.

\subsection{Multiparameter estimation}

In multiparameter estimation, the
quantum Fisher information matrix $\mathcal{F}$ is also defined via
the SLD operators, i.e.,
\begin{equation}
\mathcal{F}_{ij}:=\frac{1}{2}\langle\{L_{\theta_i},L_{\theta_j}\}\rangle,
\end{equation}
where $L_{\theta_{i(j)}}$ is the SLD operator for parameter $\theta_{i(j)}$.
For any state satisfying Eq.~(\ref{eq:rho_form}), $\mathcal{F}_{ij}$
can be expressed by
\begin{equation}
\mathcal{F}_{ij} = \frac{1}{\alpha^{2}}\Big[\alpha\mathrm{Tr}
\left\{ \partial_{i}\rho,\partial_{j}\rho\right\} +\partial_{i}\beta
\left(\partial_{j}\beta\right)\mathrm{Tr}\rho^{-1} 
-\left(\!M-\frac{1}{2}\!\right)\!\!\left(\partial_{i}\alpha\partial_{j}\beta
 +\partial_{j}\alpha\partial_{i}\beta\right)\!\!\Big].
\end{equation}
Here $\partial_{i(j)}$ represents the partial derivative on $\theta_{i(j)}$. 
Obviously, the diagonal element of $\mathcal{F}$ reduces to the form in
Eq.~(\ref{eq:F}). For the cases that $\alpha$ is constant and $\beta$
is dependent on the parameters, $\mathcal{F}_{ij}$ is in the form
\begin{equation}
\mathcal{F}_{ij}=\frac{1}{\alpha^{2}}\left[\alpha\mathrm{Tr}
\left\{ \partial_{i}\rho,\partial_{j}\rho\right\} +\partial_{i}\beta
\left(\partial_{j}\beta\right)\mathrm{Tr}\rho^{-1}\right].
\end{equation}
Especially, for a two-level state, this equation reduces to
\begin{equation}
\mathcal{F}_{\mathrm{q},ij}=\mathrm{Tr}\left(\left\{ \partial_{i}\rho,\partial_{j}\rho\right\} \right) 
+\frac{\partial_{i}\mathcal{P}\partial_{j}\mathcal{P}}
{2(1-\mathcal{P})}.
\end{equation}
This is the general basis-independent expression of quantum Fisher
information matrix for any two-level state.

\section{Conclusion}
In summary, we first reviewed the Lyapunov representation
of the SLD operator and showed that this representation is
available for both full and non-full rank density matrices. Furthermore,
based on the Lyapunov representation, we gave a further method for 
the calculation of SLD operator. This method is particularly useful
for those states whose anti-commutators with their partial
derivatives exhibits periodic properties.

As an application of the given method, we discussed a class of states,
which have a linear relation with their squares. The corresponding
analytical expressions of the SLD operator and QFI
are provided via the method. Furthermore, we discussed the depolarizing channel   
scenario of these states and extend our discussion to the block diagonal states. 
For multiparameter estimation, the quantum
Fisher information matrix is also analytically given.

The calculation of SLD operator is an important topic in theoretical
quantum metrology. We hope this work may draw attention in the
community to studying more methods to obtain the SLD operators for
various scenarios.

\ack
The authors thank Y. Yao and X. Xiao for helpful discussions. This work was 
supported by the NFRPC through Grant No.~2012CB921602 and the NSFC 
through Grants No.~11475146.

\appendix

\section{SLD for states commuting with their partial derivative}

In the following we give the detailed calculation of the SLD operator
for the states commuting with their partial derivative. From the equation
\begin{equation}
L=-2\lim_{s\rightarrow\infty}\sum_{n=0}^{\infty}\sum_{m=0}^{n}
\frac{\left(-s\right)^{n+1}}{(n+1)!}C_{n}^{m}\rho^{m}
\left(\partial_{\theta}\rho\right)\rho^{n-m},
\end{equation}
one can see that when $\rho$ commutes with $\partial_{\theta}\rho$, this equation
can be rewritten into
\begin{equation}
L=-\lim_{s\rightarrow\infty}\sum_{n=0}^{\infty}\frac{\left(-2s\right)^{n+1}}{(n+1)!}
\rho^{n}\left(\partial_{\theta}\rho\right).
\end{equation}
Remind that the spectral decomposition of density matrix is in the form
\begin{equation}
\rho=\sum_{i=1}^{M}p_{i}|\psi_{i}\rangle\langle\psi_{i}|,
\end{equation}
where $p_{i}$ and $|\psi_{i}\rangle$ are $i$th eigenvalue and eigenstate
of $\rho$, respectively. $M$ is the dimension of $\rho$'s support.
In this representation, the SLD operator is
\begin{eqnarray}
L &=& -\lim_{s\rightarrow\infty}\sum_{n=0}^{\infty}
\frac{\left(-2s\right)^{n+1}}{(n+1)!}
\sum_{i=1}^{M}p_{i}^{n}|\psi_{i}\rangle
\langle\psi_{i}|\partial_{\theta}\rho
\nonumber \\
&=& -\sum_{i=1}^{M}\frac{1}{p_{i}}\lim_{s\rightarrow\infty}
\left(e^{-2sp_{i}}-1\right)
|\psi_{i}\rangle\langle\psi_{i}|\partial_{\theta}\rho  \nonumber \\
&=& \sum_{i=1}^{M}\frac{1}{p_{i}}|\psi_{i}\rangle\langle\psi_{i}|
\partial_{\theta}\rho.
\end{eqnarray}
It is known that $\sum_{i=1}^{M}p_{i}^{-1}|\psi_{i}\rangle\langle\psi_{i}|$
is defined as the inverse matrix of $\rho$ on the support.
Thus, the SLD operator can be finally expressed by
\begin{equation}
L=\rho^{-1}\partial_{\theta}\rho,
\end{equation}
where $\rho^{-1}$ is the inverse matrix of $\rho$ on the support.

\section{Detailed calculation for the application}

In the application, for any state satisfying the equation
\begin{equation}
\rho^{2}=\alpha\rho-\beta,\label{eq:rho_form-1}
\end{equation}
one can obtain the following relation
\begin{equation}
\rho^{\mathrm{o}}\left(\partial_{\theta}\rho\right)=\alpha\partial_{\theta}\rho
+\left(\partial_{\theta}\alpha\right)\rho-\partial_{\theta}\beta.\label{eq:material_rho}
\end{equation}
Based on this equation and Eq.~(\ref{eq:rho_form-1}), the $n$th
order term is in the form
\begin{equation}
\left(\rho^{\mathrm{o}}\right)^{n}\partial_{\theta}\rho=\alpha^{n}\partial_{\theta}\rho
+\left(\rho\partial_{\theta}\alpha-\partial_{\theta}\beta\right)\alpha^{n-1}\sum_{m=0}^{n-1}
\left(\frac{2\rho}{\alpha}\right)^{m}.
\end{equation}
Submitting this equation into the equation
\begin{equation}
L=-2\lim_{s\rightarrow\infty}\sum_{n=0}^{\infty}\frac{\left(-s\right)^{n+1}}{(n+1)!}
\left(\rho^{\mathrm{o}}\right)^{n}\partial_{\theta}\rho
\end{equation}
and since $\alpha>0$, the SLD operator can be expressed by
\begin{eqnarray}
L & = & \frac{2}{\alpha}\partial_{\theta}\rho-2\left(\rho\partial_{\theta}\alpha-\partial_{\theta}
\beta\right)\times\nonumber \\
&  & \lim_{s\rightarrow\infty}\sum_{n=0}^{\infty}\frac{\left(-s\right)^{n+1}}{(n+1)!}
\alpha^{n-1}\sum_{m=0}^{n-1}\left(\frac{2\rho}{\alpha}\right)^{m}.\label{eq:L_good}
\end{eqnarray}
Utilizing the spectral decomposition of the density matrix, the term
\begin{eqnarray}
& & \lim_{s\rightarrow\infty}\sum_{n=0}^{\infty}\frac{\left(-s\right)^{n+1}}{(n+1)!}
\alpha^{n-1}\sum_{m=0}^{n-1}\left(\frac{2\rho}{\alpha}\right)^{m}\nonumber \\
&=&\!\! \lim_{s\rightarrow\infty}\sum_{i=1}^{M}\sum_{n=0}^{\infty}\frac{\left(-s\right)^{n+1}}
{(n+1)!}\alpha^{n-1}\sum_{m=0}^{n-1}\left(\frac{2}{\alpha}\right)^{m}p_{i}^{m}
|\psi_{i}\rangle\langle\psi_{i}|\nonumber \\
&=&\!\! \lim_{s\rightarrow\infty}\!\sum_{i=1}^{M}\!\frac{\alpha}{\alpha-2p_{i}}\!
\sum_{n=0}^{\infty}\!\frac{\left(-s\right)^{n+1}}{(n+1)!}\alpha^{n-1}\!\!
\left[\!1-\left(\!\frac{2p_{i}}{\alpha}\!\right)^{\!n}\!\right]\!\!|\psi_{i}
\rangle\langle\psi_{i}|
\nonumber \\
&=&\!\!\lim_{s\rightarrow\infty}\!\sum_{i=1}^{M}\!\frac{1}{\alpha\!-\!2p_{i}}
\!\left[\frac{1}{\alpha}\!\left(\!e^{-s\alpha}-1\!\right)\!-\!\frac{1}{2p_{i}}\!
\left(\!e^{-2sp_{i}}-1\!\right)\!\right]\!\!|\psi_{i}\rangle\langle\psi_{i}|, \nonumber
\end{eqnarray}
where the equality $\sum_{m=0}^{n-1}x^{m}=(1-x^{n})/(1-x)$ has been
applied. Since all $p_{i}$ here are larger than zero, above limitation
reduces to the form
\begin{equation}
\frac{1}{2\alpha}\sum_{i=1}^{M}p_{i}^{-1}|\psi_{i}\rangle\langle\psi_{i}|
=\frac{1}{2\alpha}\rho^{-1},
\end{equation}
where $\rho^{-1}$ is the inverse matrix of $\rho$ on the support.
Finally, the SLD operator for any state satisfying Eq.~(\ref{eq:rho_form-1})
can be expressed by
\begin{equation}
L=\frac{1}{\alpha}\left[2\partial_{\theta}\rho+\left(\partial_{\theta}\beta\right)
\rho^{-1}-\partial_{\theta}\alpha \bm{I}_{\mathrm{M}}\right],\label{eq:material_L}
\end{equation}
where  $\bm{I}_{\mathrm{M}}$ is the identity matrix in the support of  $\rho$.

\subsection{Calculation of QFI} 

Based on Eq.~(\ref{eq:material_L}), the quantum Fisher information
$F=\langle L^{2}\rangle$ can be directly calculated as
\begin{eqnarray}
F &=& \frac{1}{\alpha^{2}}\Big[4\langle\left(\partial_{\theta}\rho\right)^{2}\rangle
-4\left(\partial_{\theta}\alpha\right)\langle\partial_{\theta}\rho\rangle\nonumber 
+\left(\partial_{\theta}\beta\right)^{2}\mathrm{Tr}\left(\rho^{-1}\right) \\
& & -2M\left(\partial_{\theta}\alpha\right)\left(\partial_{\theta}\beta\right)
+\left(\partial_{\theta}\alpha\right)^{2}\Big],
\end{eqnarray}
where $\langle\{\partial_{\theta}\rho,\rho^{-1}\}\rangle=2\mathrm{Tr}(\partial_{\theta}\rho)=0$
has been applied. Denoting the purity of $\rho$ as $\mathcal{P}$,
i.e., $\mathcal{P}=\mathrm{Tr}\rho^{2}$, one can see that $\langle\partial_{\theta}
\rho\rangle=\partial_{\theta}\mathcal{P}/2$,
and
\begin{equation}
\langle\left(\partial_{\theta}\rho\right)^{2}\rangle=\frac{1}{2}\mathrm{Tr}
\left[\left(\rho^{\mathrm{o}}\partial_{\theta}\rho\right)\partial_{\theta}\rho\right].
\end{equation}
Substituting Eq.~(\ref{eq:material_rho}) into above equation, there is
\begin{equation}
\langle\left(\partial_{\theta}\rho\right)^{2}\rangle=\frac{1}{2}\alpha\mathrm{Tr}
\left(\partial_{\theta}\rho\right)^{2}+\frac{1}{4}\left(\partial_{\theta}\alpha\right)
\left(\partial_{\theta}\mathcal{P}\right).
\end{equation}
Therefore, the quantum Fisher information can be expressed by
\begin{eqnarray}
F & = & \frac{1}{\alpha^{2}}\Big[2\alpha\mathrm{Tr}\left(\partial_{\theta}\rho\right)^{2}
-\left(\partial_{\theta}\alpha\right)\left(\partial_{\theta}\mathcal{P}\right)
+\left(\partial_{\theta}\alpha\right)^{2}\nonumber \\
 &  & +\left(\partial_{\theta}\beta\right)^{2}\mathrm{Tr}\left(\rho^{-1}\right)
 -2M\left(\partial_{\theta}\alpha\right)\left(\partial_{\theta}\beta\right)\Big].
\end{eqnarray}
Moreover, since $\mathcal{P}=\alpha-\beta$ here, above equation can finally be
written as
\begin{eqnarray}
F &=& \frac{1}{\alpha^{2}}\Big[2\alpha\mathrm{Tr}\left(\partial_{\theta}\rho\right)^{2}
+\left(\partial_{\theta}\beta\right)^{2}\mathrm{Tr}\rho^{-1} \nonumber \\
& & -\left(2M-1\right)\left(\partial_{\theta}\alpha\right)
\left(\partial_{\theta}\beta\right)\Big].\label{eq:material_F}
\end{eqnarray}

\subsection{Calculation of quantum Fisher information matrix}

For the SLD operators in Eq.~(\ref{eq:material_L}), the element
of quantum Fisher information matrix is
\begin{eqnarray}
\mathcal{F}_{ij}\!\! &=& \!\! 2\mathrm{Tr}\left(\rho\left\{ \partial_{i}\rho,\partial_{j}\rho\right\} \right)-\partial_{j}\alpha\partial_{i}\mathcal{P}-\partial_{i}\alpha\partial_{j}
\mathcal{P}+\partial_{i}\alpha\partial_{j}\alpha \nonumber \\
& & +\left(\partial_{i}\beta\right)\left(\partial_{j}\beta\right)
\mathrm{Tr}\rho^{-1}-M\left(\partial_{j}\alpha\partial_{i}\beta
+\partial_{i}\alpha\partial_{j}\beta\right). \nonumber
\end{eqnarray}
Since the first term in above equation can be rewritten into
\begin{eqnarray}
& &2\mathrm{Tr}\left(\rho\left\{\partial_{i}\rho,\partial_{j}\rho\right\} \right) \nonumber \\
&=& \frac{1}{2}\left(\partial_{i}\alpha\partial_{j}\mathcal{P}+\partial_{j}
\alpha\partial_{i}\mathcal{P}\right)
+\alpha\mathrm{Tr}(\left\{\partial_{i}\rho,\partial_{j}\rho\right\}),
\nonumber
\end{eqnarray}
and $\partial_{i}\mathcal{P}=\partial_{i}\alpha-\partial_{i}\beta$,
$\mathcal{F}_{ij}$ can then be simplified as
\begin{eqnarray}
\mathcal{F}_{ij} & = & \frac{1}{\alpha^{2}}\Big[\alpha\mathrm{Tr}\left\{ \partial_{i}\rho,\partial_{j}\rho\right\} +\partial_{i}\beta\left(\partial_{j}\beta\right)\mathrm{Tr}\rho^{-1}
\nonumber \\
& & -\left(M-\frac{1}{2}\right)\left(\partial_{i}\alpha\partial_{j}\beta
+\partial_{j}\alpha\partial_{i}\beta\right)\Big].
\end{eqnarray}
When $i=j$, above equation reduces to Eq.~(\ref{eq:material_F}).

\end{document}